\documentclass{PoS}

\usepackage{epstopdf}

\newcommand{\be}{\begin{equation}}
\newcommand{\ee}{\end{equation}}
\newcommand{\ba}{\begin{eqnarray}}
\newcommand{\ea}{\end{eqnarray}}

\newcommand{\fig}{Fig.~}

\def\lsi{\raise0.3ex\hbox{$<$\kern-0.75em\raise-1.1ex\hbox{$\sim$}}}
\def\gsi{\raise0.3ex\hbox{$>$\kern-0.75em\raise-1.1ex\hbox{$\sim$}}}
\newcommand{\lsim}{\mathop{\lsi}}

\title{
{\vspace{-25mm} \normalsize\hfill{\small MS-TP-09-20}}\\[25mm]
Towards a determination of the chiral critical surface of QCD}

\ShortTitle{Chiral critical surface of QCD}

\author{\speaker{Owe Philipsen}\thanks{In collaboration with Ph.~de Forcrand}\\
        Westf\"alische Wilhelms-Universit\"at M\"unster, 48149 M\"unster, Germany\\
        E-mail: \email{ophil@uni-muenster.de}}

\abstract{
The chiral critical surface is a surface of
second order phase transitions bounding the region of first order
chiral phase transitions for small quark masses in the $\{m_{u,d}, m_s,\mu\}$ parameter space.
The potential critical endpoint of the QCD ($T,\mu$)-phase diagram is widely
expected to be part of this surface.
Since for $\mu=0$ with physical quark masses QCD is known to exhibit an analytic crossover,
this expectation requires the region of chiral transitions to expand with $\mu$ for a chiral 
critical endpoint to exist.
Instead, on coarse $N_t=4$ lattices, we find the area of chiral transitions to shrink with
$\mu$, which excludes a chiral critical point for QCD at moderate
chemical potentials $\mu_B < 500$ MeV.
First results on finer $N_t=6$ lattices indicate a curvature of the critical surface 
consistent with zero and unchanged conclusions. We also comment on the interplay of phase diagrams between the $N_f=2$ and $N_f=2+1$ theories and its consequences for physical QCD.
}

\FullConference{5th International Workshop on Critical Point and Onset of Deconfinement - CPOD 2009,\\
		 June 08 - 12 2009\\
		 Brookhaven National Laboratory, Long Island, New York, USA}

\begin{document}

\section{Introduction}

The QCD phase diagram has been the subject of intense research over the last ten years. 
Based on asymptotic freedom, one expects at least 
three different forms of nuclear matter: confined hadronic matter (low $\mu_B,T$), 
quark gluon plasma (high $T$) and colour-superconducting matter (high $\mu_B$, low $T$). 
Whether and where these regions are separated
by true phase transitions has to be determined by first principle calculations and experiments.
Since QCD is strongly coupled on scales of nuclear matter, Monte Carlo simulations
of lattice QCD are presently the only viable approach.

Unfortunately, the so-called sign problem prohibits straightforward
simulations at finite bary\-on density, thus our expectations for the QCD phase diagram are largely
founded on model calculations. Since 2001, several ways have been designed to circumvent the sign problem in an approximate way, all of them valid for $\mu/T\lsim1$ only \cite{oprev,csrev}. 
Within this range, those methods give quantitatively agreeing results for, e.g., the calculation of 
$T_c(\mu)$ \cite{slavo}. Many phenomenologically interesting quantities like 
screening masses, the thermodynamical pressure, quark number susceptibilities etc.~are thus
theoretically controlled at moderate quark densities. On the other hand,
because of the intricate and costly finite size scaling analyses involved, determining the order of
the QCD phase transition, and hence the existence of a chiral critical point, is a much harder task.
Here we discuss the problem of determining the order of the finite temperature phase transition 
by lattice simulations in the extended parameter space 
$\{m_{u,d},m_s,T,\mu\}$. 

\section{The chiral critical line at $\mu=0$}

\begin{figure}[t]
\begin{center}
\includegraphics[width=0.4\textwidth]{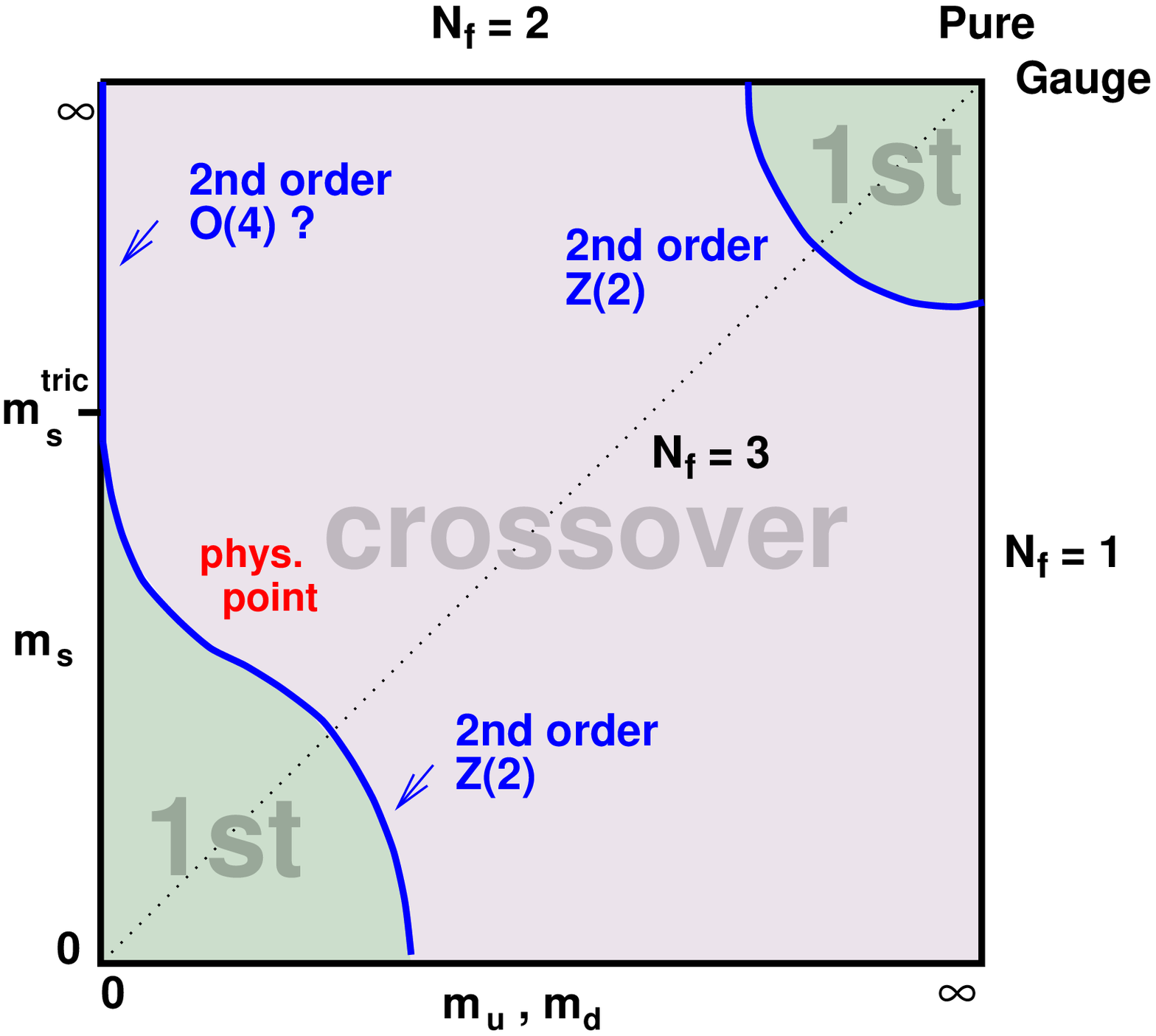}
\includegraphics[width=0.5\textwidth]{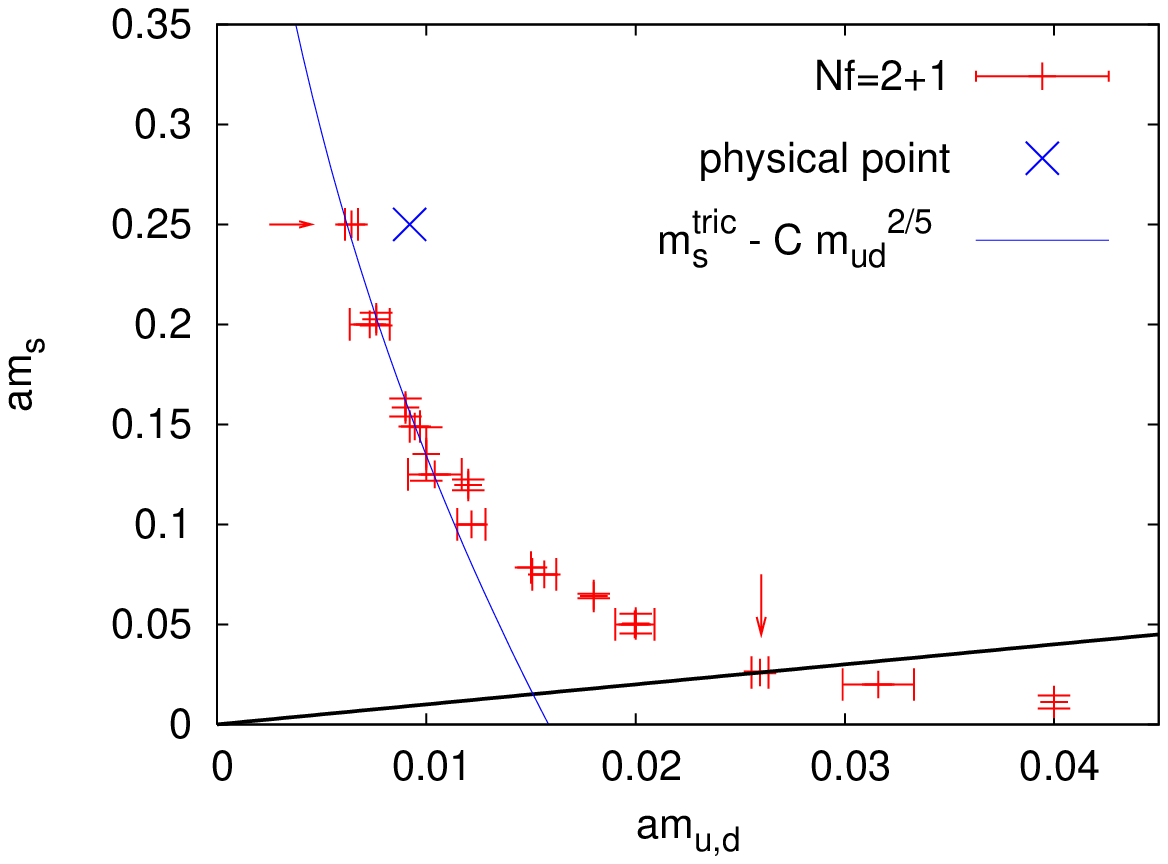}
\end{center}
\caption{\label{schem} Left: Schematic phase transition behaviour of $N_f=2+1$
QCD for different choices of quark masses $(m_{u,d},m_s)$ at
$\mu=0$. Right: Measured chiral critical line on an $N_t=4$ lattice with staggered fermions \cite{fp3}.
}
\end{figure}

The order of the QCD finite temperature phase transition as a function of quark masses 
is depicted in \fig\ref{schem}, for $\mu=0$ (left).
In the limits of zero and infinite quark masses (lower left and upper 
right corners), order parameters corresponding to the breaking of  
the global chiral and centre symmetries, respectively,  
can be defined, and one numerically finds first order phase
transitions at small and large quark masses at some finite
temperatures $T_c(m_{u,d},m_s)$. On the other hand, one observes an analytic crossover at
intermediate quark masses, with second order boundary lines separating these
regions. Both lines have been shown to belong to the $Z(2)$ universality class
of the 3d Ising model \cite{kls,fp2,kim1}. Since the line on the lower left marks the boundary
of the quark mass region featuring a chiral phase transition, it is referred to as chiral critical line.

However, the nature of the $N_f=2$ chiral transition is far from being settled.
Wilson fermions appear to see O(4) scaling \cite{wil}, 
while staggered actions are inconsistent with O(4) and O(2) (for the discretised theory) 
\cite{s2}. A recent finite size scaling analysis using staggered fermions
with unprecedented lattice sizes was performed in \cite{dig}. 
Again, these data appear inconsistent with O(4)/O(2), and the authors
conclude a first order transition to be a possibility.
A different conclusion was reached in \cite{ksnf2}, in which $\chi$QCD was 
investigated numerically.
This is a staggered action modified by an irrelevant term  such as to allow 
simulations in the chiral limit. 
The authors find their data compatible with those of an 
$O(2)$ spin model on moderate to small volumes,
which would indicate large finite volume effects in the other simulations.
Finally, from universality of chiral models it is known that the order of the chiral transition
is related to the strength of the $U_A(1)$ anomaly \cite{piwi}.
In a model constructed to have the right symmetry with a tunable anomaly strength, it has recently been demonstrated non-perturbatively that both scenarios are possible,
with a strong anomaly required for the chiral phase transition to be second order \cite{chi}.
Should the chiral transition turn out 
to be first order, the likely modification of  \fig\ref{schem} (left) 
would be the disappearance of the 
tricritical point, with the chiral critical line intersecting the $N_f=2$ axis at 
some finite $m_{u,d}$ and being Z(2) all the way. 

\begin{figure}[t]
\begin{center}
\includegraphics[width=0.48\textwidth]{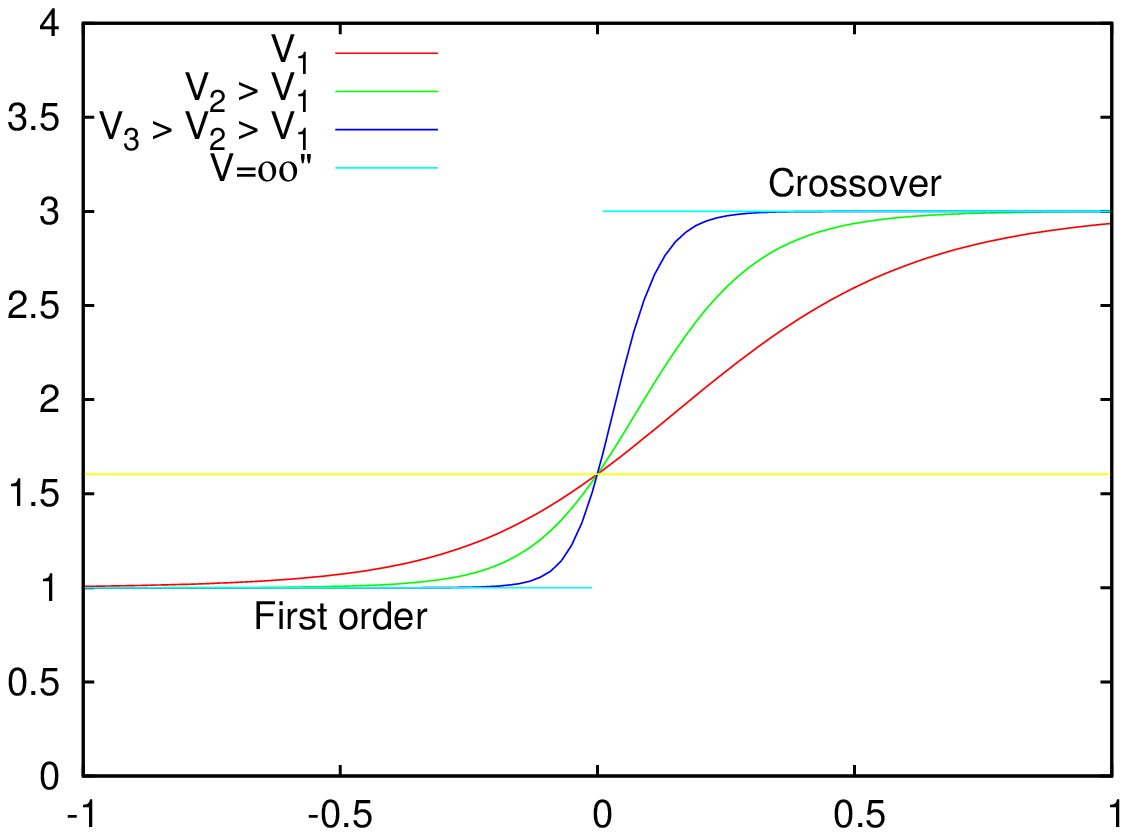}
\includegraphics[width=0.5\textwidth]{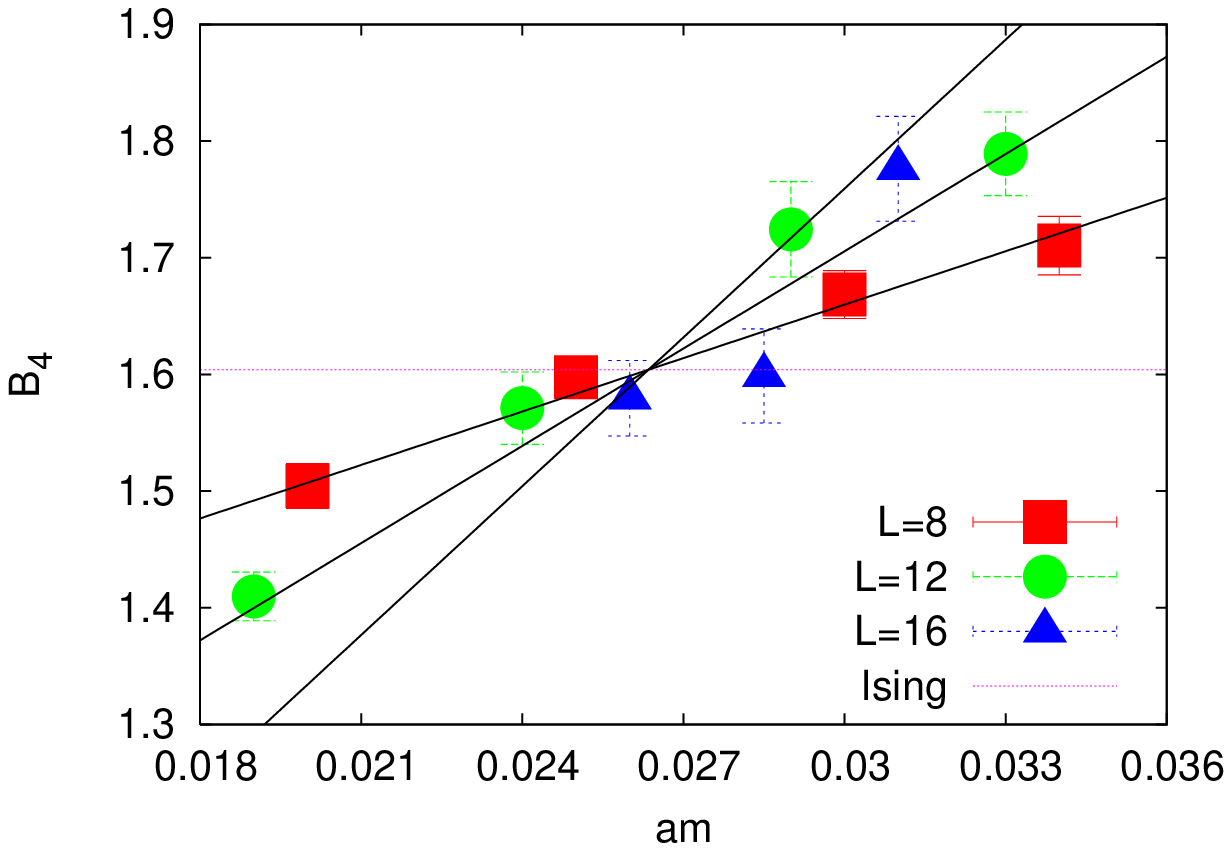}
\end{center}
\caption{\label{b4schem} Left: Schematic behaviour of the Binder cumulant as a function of 
the $N_f=3$ quark mass at finite and infinite volume. Right: Data obtained on $N_t=4$ lattices with staggered fermions \cite{fp3}.
}
\end{figure}
A convenient observable to locate and identify the second order boundary lines
is the Binder cumulant
\be
B_4(X) \equiv \langle (X - \langle X \rangle)^4 \rangle / \langle (X - \langle X \rangle)^2 \rangle^2, \quad
X \!=\! \bar\psi \psi\;.
\ee
It has to be evaluated at the (pseudo-)critical coupling $\beta_c(m,\mu)$, i.e.~on the phase boundary defined by a vanishing third moment of the fluctuation, 
$\langle (X-\langle X\rangle)^3\rangle|_{\beta_c}=0$.
In the infinite volume limit, $B_4\rightarrow 1,3$ for a first order transition or crossover, respectively.
At the second order transition, $B_4\rightarrow 1.604$ dictated by the 
$3d$ Ising universality class to which the chiral critical line belongs.
On finite lattices this step function gets smeared out to an analytic function, 
as shown in \fig\ref{b4schem} (left).
The intersection points of different volumes serve as estimators of the critical value, $B_4(\beta_c(m_c),m_c)$.
A scan of $B_4(\beta_c(m),m)$ in the $N_f=3$ theory is shown in \fig\ref{b4schem} (right).  
Also shown is a simultaneous fit of all three volumes to a Taylor expansion of the cumulant
around the critical point and exploiting the theoretically known behaviour under finite size scaling,
\be
B_4(m,L)=1.604+bL^{1/\nu}(m-m_c)+\ldots
\ee
Such a fit allows to check whether the chosen volumes are large enough to be consistent with finite
size scaling, as well as extracting the universality class from the values of $B_4$ at the intersection point
and the scaling exponent, $\nu=0.63$ in the case of 3d Ising.

Following this recipe by fixing $m_s$ and then scanning in $m_{u,d}$, the chiral critical line has recently been mapped out on $N_t=4$ lattices \cite{fp3}, \fig\ref{schem} (right).
In agreement with expectations, the critical line steepens when approaching the chiral limit. Assuming 
the $N_f=2$ chiral transition to be in the O(4) universality class implies a tricritical point on the $m_s$-axis, \fig\ref{schem} (left). The data are consistent with tricritical scaling \cite{derivs} of the critical line with $m_{u,d}$ and we estimate $m_s^{tric}\sim 2.8 T_c$. However, this value is extremely cut-off sensitive and likely smaller in the continuum, cf.~Sec.\ref{sec:nt6}.

\begin{figure}[t]
\includegraphics[width=0.5\textwidth]{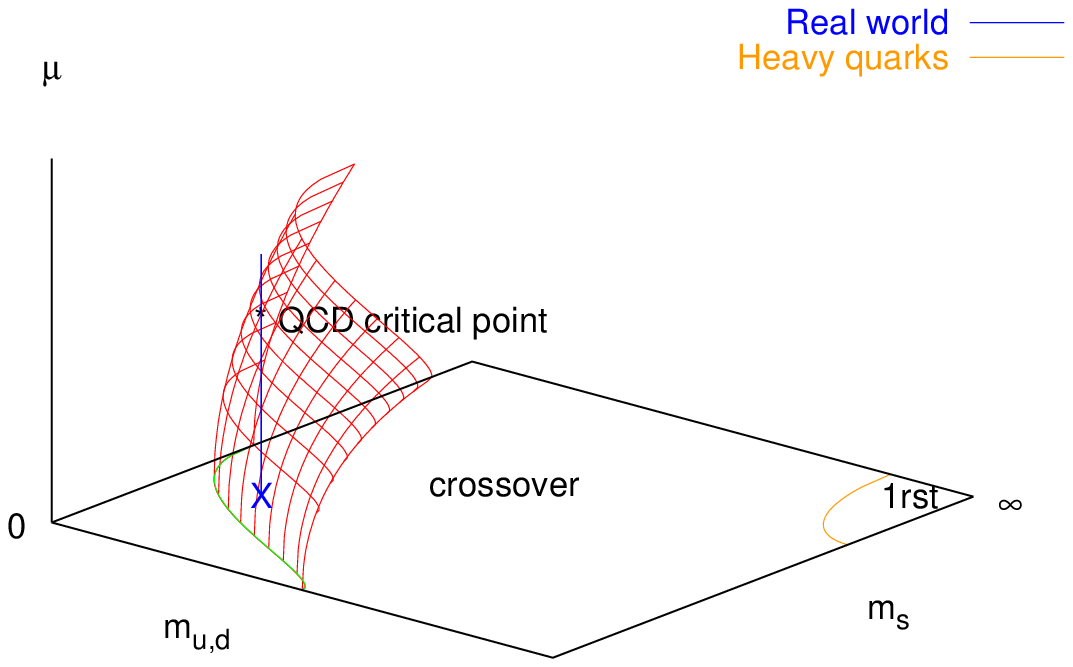}
\includegraphics[width=0.5\textwidth]{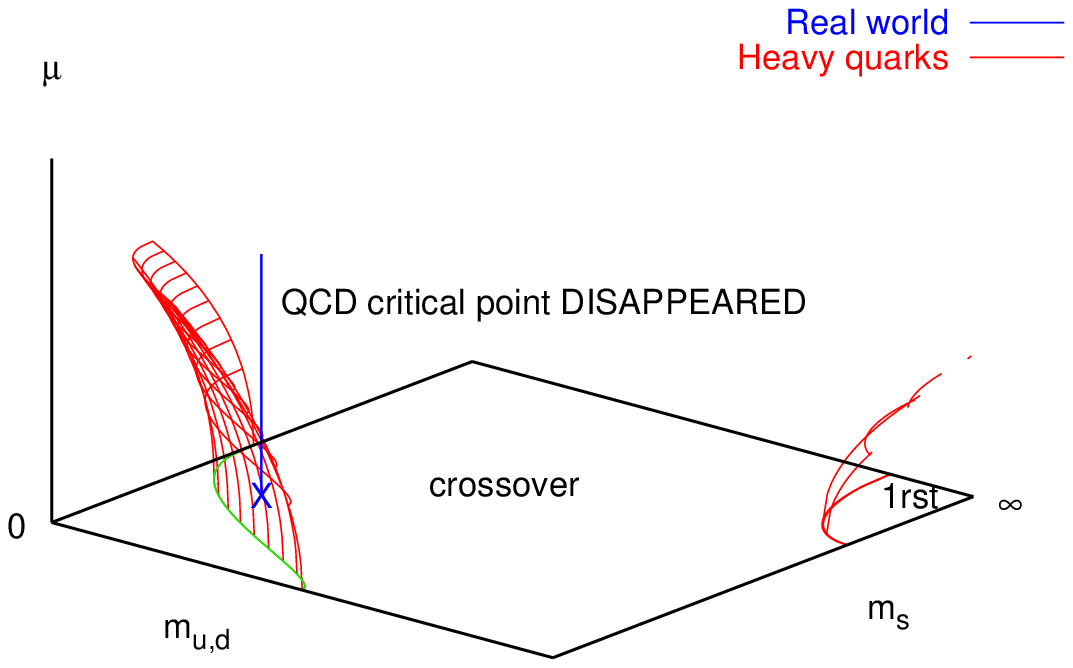}
\caption[]{Critical surface swept by the chiral 
critical line as $\mu$ is turned on. Depending on the curvature, a QCD chiral critical
point is present or absent. 
For heavy quarks the curvature has been determined \cite{kim1} and the first order 
region shrinks with $\mu$.
}
\label{schem3}
\end{figure}

\section{The chiral critical surface}

When a chemical potential is switched on, the chiral critical line sweeps out a surface, as shown
in \fig\ref{schem3}. According to standard expectations in the literature \cite{derivs},
for small $m_{u,d}$, the critical line should 
continuously shift with $\mu$ to larger quark masses until it passes through the physical point at $\mu_E$, corresponding to the endpoint in the QCD $(T,\mu)$ phase diagram. 
This is depicted in \fig\ref{schem3} (left), where the critical point is part of 
the chiral critical surface. However, it is also
possible for the chiral critical surface to bend towards smaller quark masses, \fig\ref{schem3} (right),
in which case there would be no chiral critical point or phase transition 
at moderate densities. For definiteness, let us consider three degenerate quarks, 
represented by the diagonal in the quark mass plane.
The critical quark mass corresponding to the boundary point has an expansion
\be
\frac{m_c(\mu)}{m_c(0)}=1+\sum_{k=1}c_k \left(\frac{\mu}{\pi T}\right)^{2k}\,.
\ee
A strategy to learn about the chiral critical surface is now to tune the quark mass to $m_c(0)$ and evaluate
the leading coefficients of this expansion. In particular, the sign of $c_1$ will tell us which of the scenarios
in \fig\ref{schem} is realised. 

\begin{figure}[t!]
\includegraphics[width=0.5\textwidth]{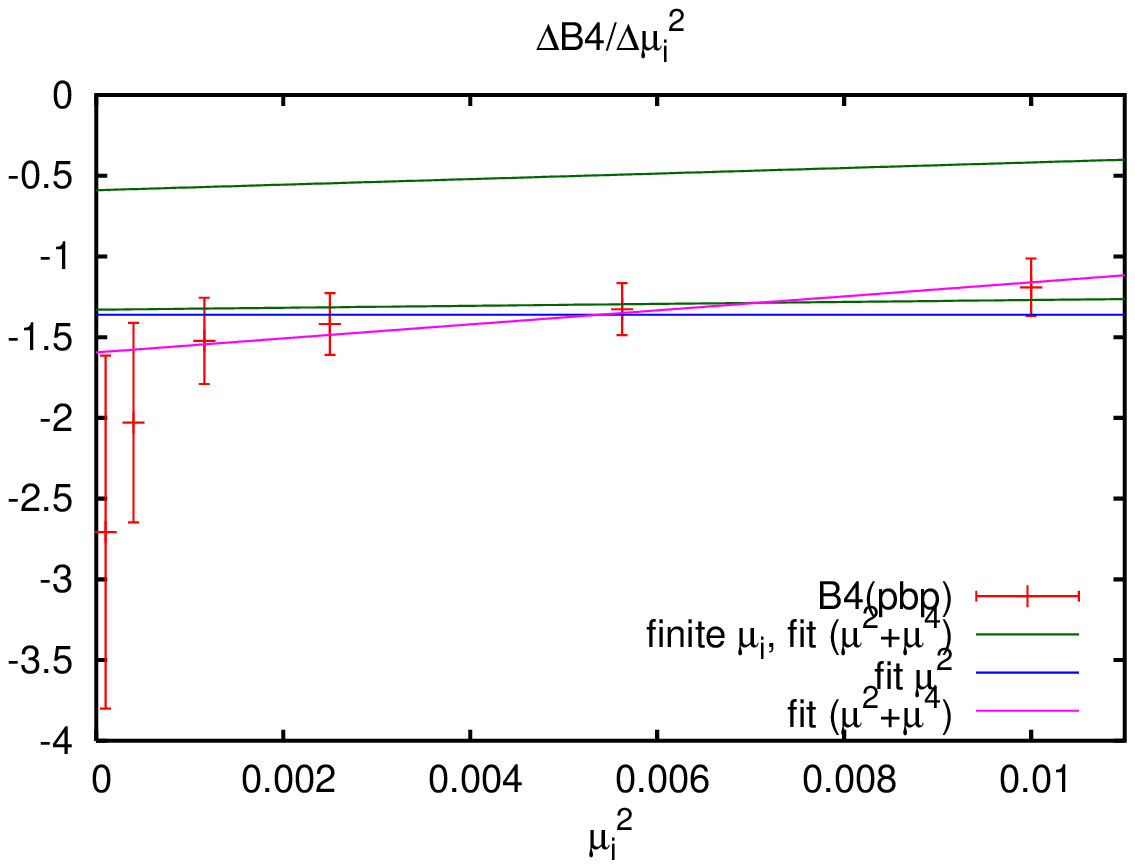}
\includegraphics[width=0.5\textwidth]{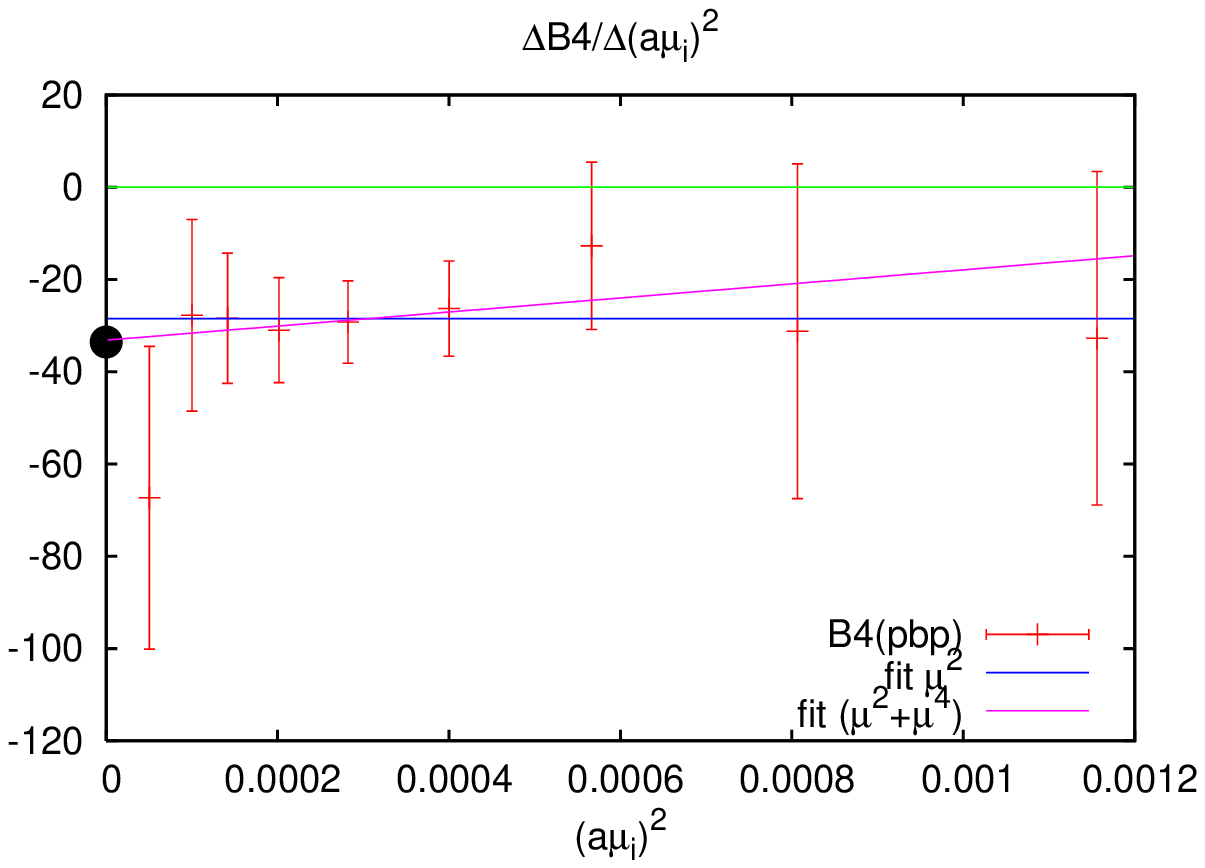}
\caption[]{$\mu^2$-dependence of the Binder cumulant on the chiral critical line for $N_f=3$ (left) \cite{fp4}
and $N_f=2+1$ with physical strange quark mass (right), both for $N_t=4$.}
\label{db4}
\end{figure}

The curvature of the critical surface in lattice units is directly related to the behaviour of the Binder cumulant via the chain rule,
\be
\frac{dam_c}{d(a\mu)^2}=-\frac{\partial B_4}{\partial (a\mu)^2}
\left(\frac{\partial B_4}{\partial am}\right)^{-1}\,.
\ee
While the second factor is sizeable and easy to evaluate, the $\mu$-dependence
of the cumulant is excessively weak and requires enormous statistics to extract. In order to guard
against systematic errors, this derivative has been evaluated in two independent ways.
One is to fit the corresponding Taylor series of $B_4$ in powers of $\mu/T$ to data generated at 
imaginary chemical potential \cite{fp3, fp4}, the other to compute the derivative directly and without
fitting via the finite difference quotient \cite{fp4},
\be
\frac{\partial B_4}{\partial (a\mu)^2}=\lim_{(a\mu)^2\rightarrow 0}\frac{B_4(a\mu)-B_4(0)}{(a\mu)^2}.
\ee 
Because the required shift in the couplings is very small,
it is adequate and safe to use the original Monte Carlo ensemble 
for $am^c(0),\mu=0$ and reweight the results by the standard 
Ferrenberg-Swendsen method. 
Moreover, by reweighting to imaginary $\mu$
the reweighting factors remain real positive and close to 1.

The results of these two procedures 
based on 20 and 5 million trajectories on $8^3\times 4$, respectively,  are shown in \fig\ref{db4} (left).
The error band represents the first coefficient from fits to imaginary $\mu$ data, while the
data points represent the finite difference quotient extrapolated to zero. Both results are consistent,
and the slope permits and extraction of the subleading $\mu^4$ coefficient, while the combination
of all data also constrains the sign of the $\mu^6$ term. 
After continuum conversion the result for $N_f=3$ is
$c_1=-3.3(3), c_2=-47(20),c_3<0$  \cite{fp4}.
The same behaviour is found for non-degenerate quark masses. Tuning the strange quark 
mass to its physical value,  we calculated
$m^{u,d}_c(\mu)$ with $c_1= -39(8)$ and $c_2<0$, \fig\ref{db4} (right).
Hence, on coarse $N_t=4$ lattices, the region of chiral phase transitions shrinks as a real chemical potential is turned on, and there is no chiral critical point for $\mu_B\lsim 500$ MeV.
Note that one also observes a weakening of the phase transition with $\mu$ in the heavy quark case \cite{kim1}, in recent model studies of the light quark regime \cite{fuk,bow}, as well as a weakening of the transition with isospin chemical potential \cite{iso}.

\section{First steps towards the continuum, $N_t=6$ \label{sec:nt6}}

\begin{figure}[t]
\begin{center}
\includegraphics[width=0.35\textwidth]{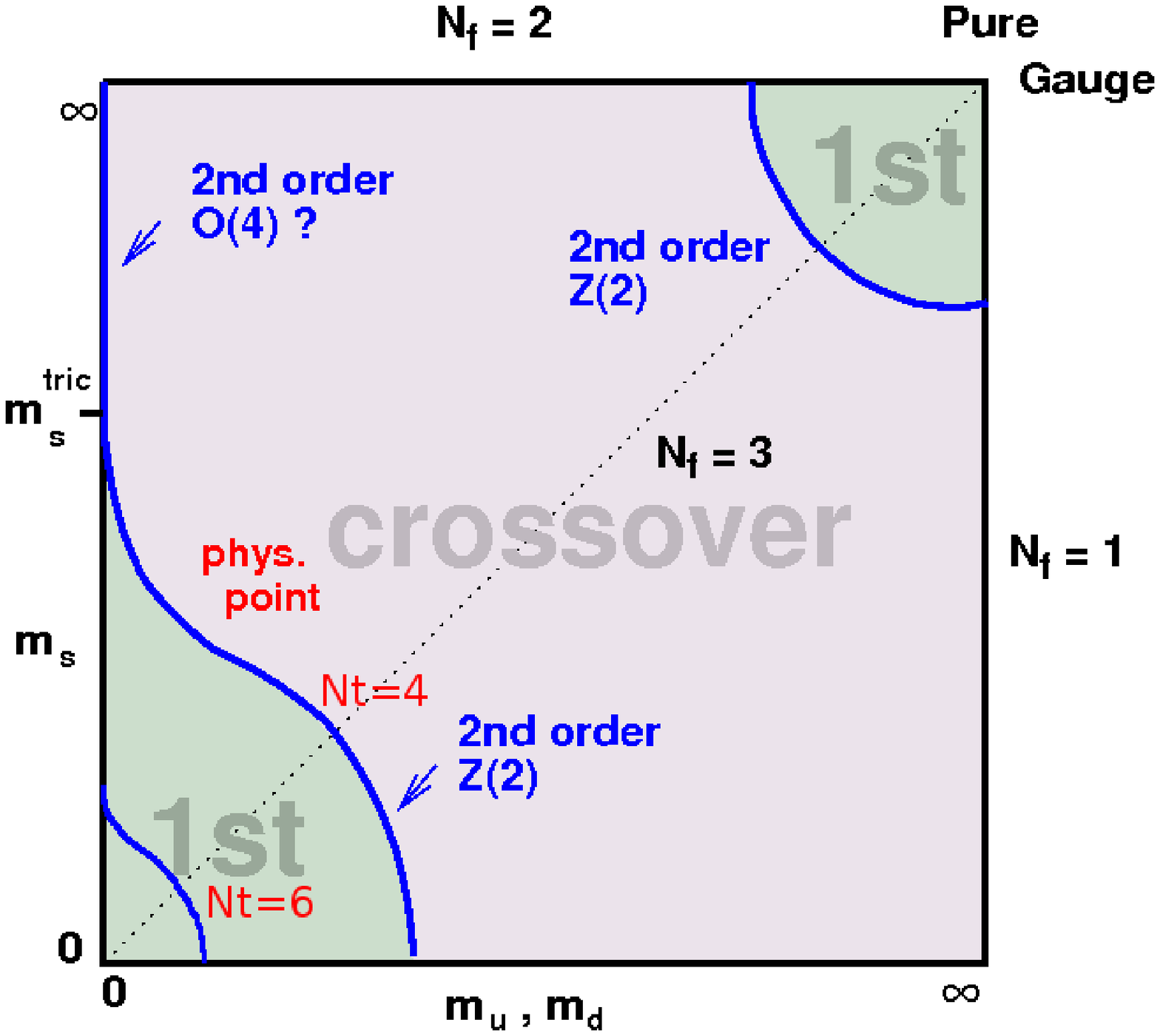}
\includegraphics[width=0.5\textwidth]{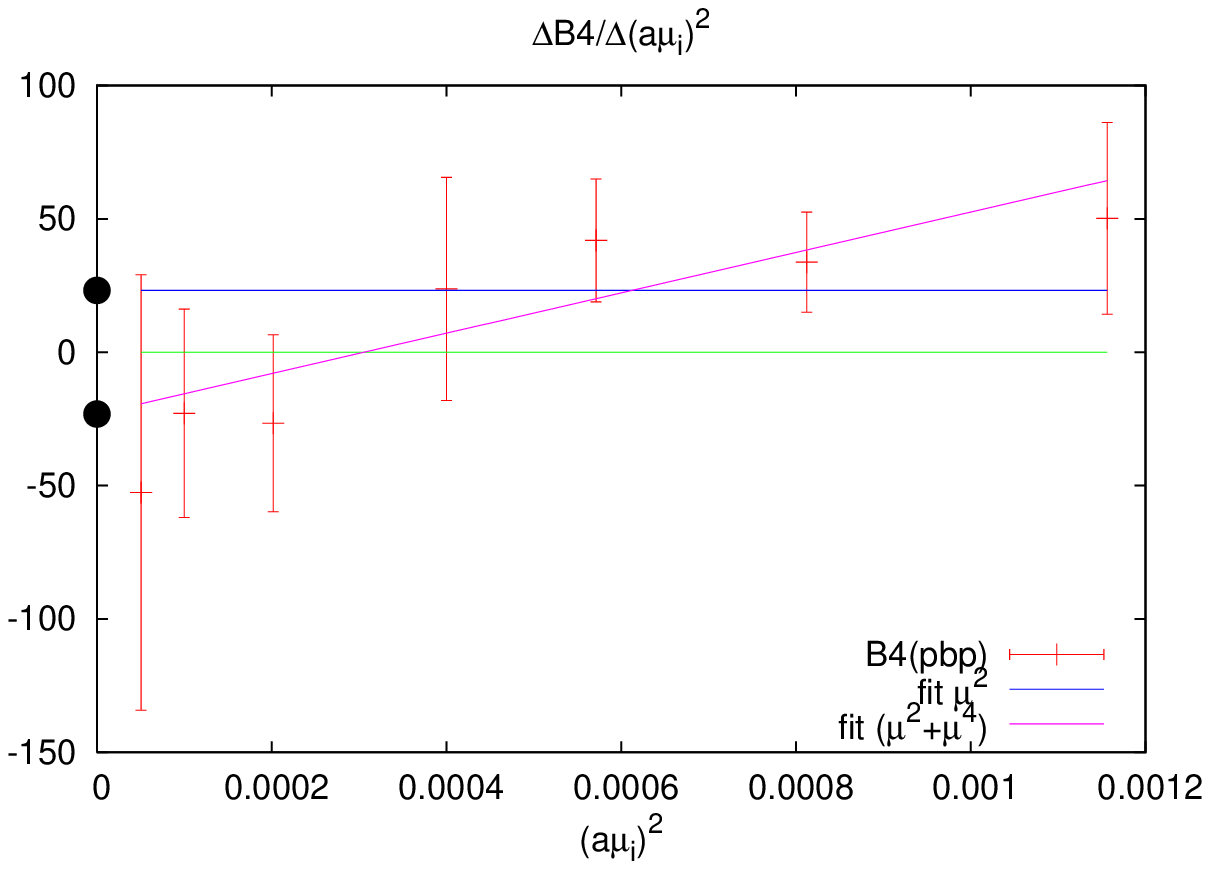}
\end{center}
\vspace*{-0.5cm}
\caption{\label{nt6} Left: The chiral critical line moves to smaller quark masses with decreasing
lattice spacing. Right: $\mu^2$-dependence of the Binder cumulant on the chiral critical line for $N_f=3,N_t=6$   }
\end{figure}

The largest uncertainty in these calculations by far is due to the coarse lattice spacing $a\sim 0.3$ fm
on $N_t=4$ lattices. First steps towards the continuum are currently being taken on $N_t=6, a\sim 0.2$ fm.  
At $\mu=0$, the chiral critical line is found to recede strongly with decreasing lattice 
spacing \cite{LAT07, fklat07}: for $N_f=3$, on the critical point $m_\pi(N_t=4)/m_\pi(N_t=6)\sim 1.8$.  
Thus, in the continuum the gap between the physical point and the chiral critical line
is much wider than on coarse lattices, as indicated in \fig\ref{nt6} (left). 
Preliminary results for the curvature of the critical surface, \fig\ref{nt6} (right), result in 
$c_1=7(14),-17(18)$ for a LO,NLO extrapolation in $\mu^2$, respectively.
Thus the sign of the curvature is not yet constrained. But even if positive, 
its absolute size appears too small to make up for the shift of the chiral critical line towards smaller quark masses, and one would again conclude for absence of a chiral critical point below $\mu_B\lsim 500$ MeV
in this approximation. Higher order terms with large coefficients would be needed to
change this picture. 

Note that on current lattices cut-off effects appear to be larger than finite density effects, hence 
definite conclusions for continuum physics cannot yet be drawn. A general finding is the 
steepness of the critical surface, making the location of a possible critical endpoint extremely quark mass sensitive, and hence difficult to determine accurately.

\section{The interplay between $N_f=2,3$ and $N_f=2+1$}

\begin{figure}[t]
\begin{center}
\includegraphics[width=0.4\textwidth]{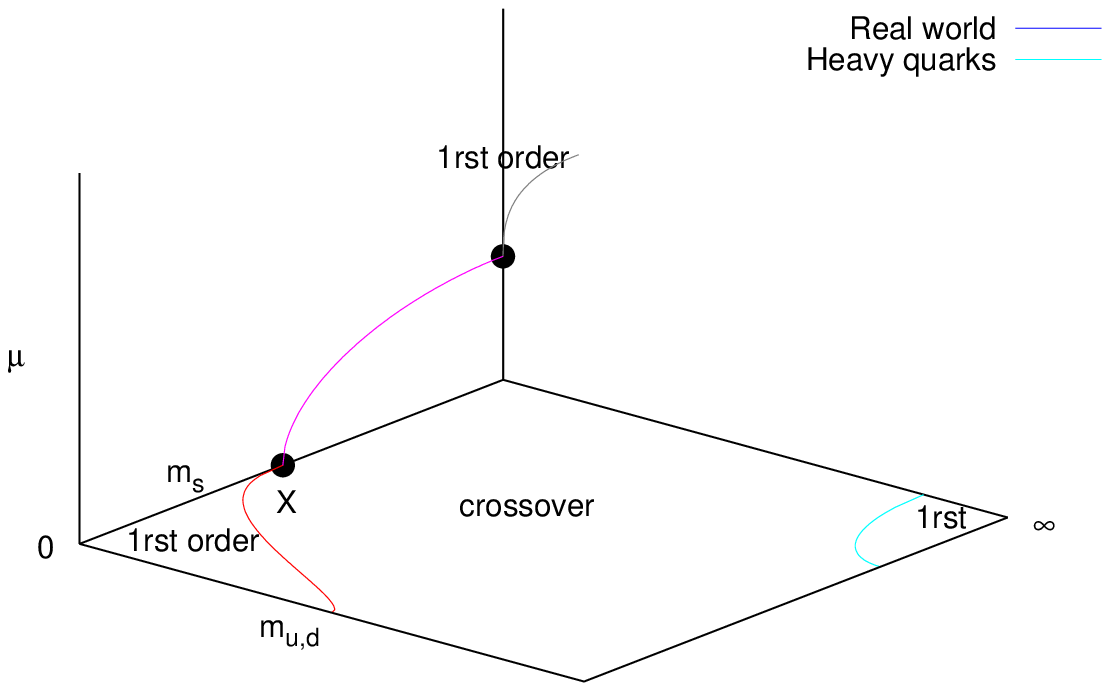}
\includegraphics[width=0.4\textwidth]{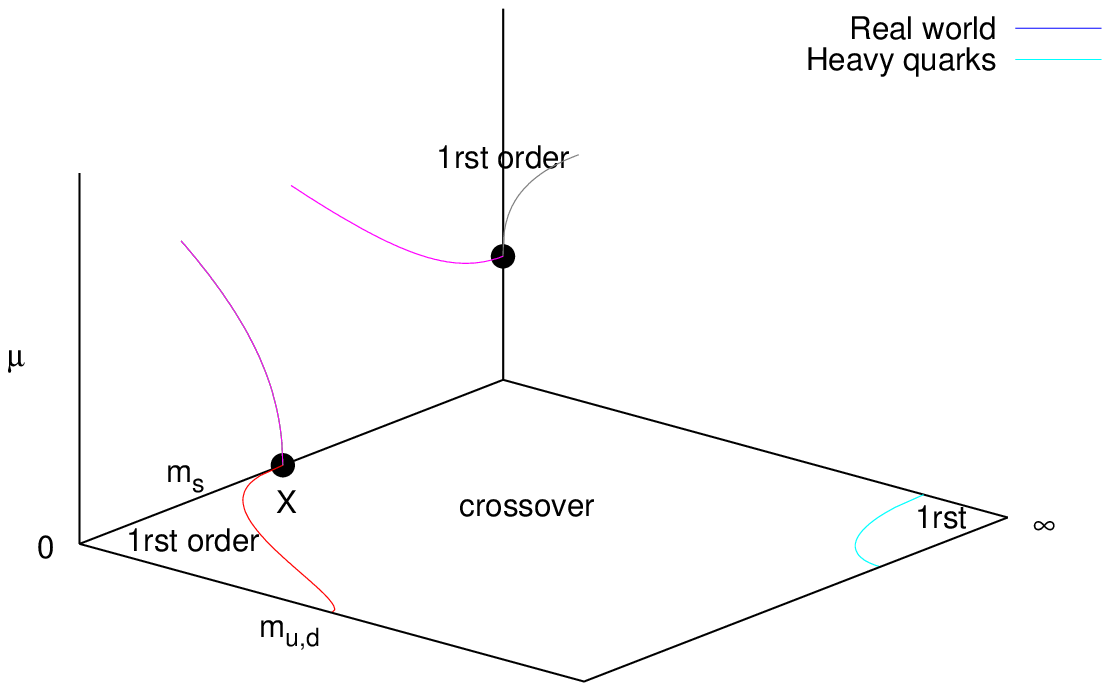}
\end{center}
\vspace*{-0.5cm}
\caption[]{\label{nf23} Two possible scenarios for connected and disconnected triple lines in the
$m_{u,d}=0$-plane.  }
\end{figure}

Let us finally comment on the importance of understanding both the $N_f=2,3$ as well as their connection
before concluding anything for physical QCD.
\fig\ref{nf23} (left) shows the scenario that is used in the literature when the argument for the expected
QCD phase diagram with a critical endpoint is made \cite{derivs}. 
The tricritical point at $\mu=0$, which we have
discussed here, is expected to be analytically connected by a tricritical line, upon varying $m_s$,
to the tricritical point at some finite $\mu$ in the two-flavour case. This picture arises plausibly
if the chiral critical surface behaves as in \fig\ref{schem3} (left). However if, as on our coarse lattices,
\fig\ref{schem3} (right) is realised, the situation might well be as in \fig\ref{nf23} (right). Even if the 
chiral $N_f=2$ theory does feature a tricritical point at finite $\mu$, it need not be connected to 
the chiral critical surface, and nothing follows for the physical point without additional information. 

Furthermore, the shrinking of the critical quark masses with diminishing lattice spacing 
makes it likely that a potential tricritical point, \fig\ref{schem}, also moves from a large value for the
strange quark mass, $m_{u,d}=0,m_s^{tric}\sim 2.8T$ on $N_t=4$ \cite{fp3}, towards smaller
values in the continuum limit. In particular, it is possible to have a situation for which
$m_{u,d}=0,m_s^{tric}<m_s^{phys}$. In this case the chiral critical surface we have been discussing here
would not be responsible for a possible critical point, regardless of its curvature, but another surface emanating
from the O(4)-chiral limit. Again, both of these scenarios depend on conclusively understanding the
situation in the $N_f=2$ chiral limit.

\section{Conclusions}

The determination of the order of the QCD finite temperature phase transition as a function
of quark chemical potential is a maximally difficult problem. Besides the sign-problem, the 
strong quark mass and flavour dependence are responsible for potentially rich structures
in the $\{m_{u,d},_s,T,\mu\}$ parameter space which are compute-expensive to disentangle due to the required intricate finite size scaling analyses in the light quark mass regime. The existing results show
that we need to control various limiting cases of QCD as well as their connection to $N_f=2+1$ 
in order to understand if there is a critical point in the QCD phase diagram, and to which critical surface
it belongs. 

\section*{Acknowledgements:}
This work is partially supported by Deutsche Forschungsgemeinschaft, project PH 158/3-1.

\end{document}